\documentstyle[twoside,fleqn,espcrc2]{article}

\title{UNIVERSALITY, SCALING
AND TRIVIALITY IN A
HIERARCHICAL SCALAR FIELD THEORY}

\author{J. J. Godina\address{
Dpt. of Physics and Astr., Univ. of Iowa, Iowa City, Iowa 52242, USA}
\address{
\it Dep. de Fis. , CINVESTAV-IPN , Ap. Post. 14-740, Mexico, D.F. 07000}
\thanks{supported by
a fellowship from CONACYT},       
Y. Meurice$^{\rm a}$, S. Niermann$^{\rm a}$, and M. B. Oktay$^{\rm a}$ 
\thanks{This work is
      supported in part by the Department of Energy
under Contract No. FG02-91ER40664. }}
\begin{document}

\begin{abstract}

Using polynomial truncations of the Fourier transform 
of the RG transformation of Dyson's hierarchical model, 
we show that it is
possible to calculate very accurately 
the renormalized quantities in the symmetric phase.
Numerical results regarding the
corrections to the scaling laws, triviality, hyperscaling, universality
and high-accuracy determinations of the
critical exponents are discussed.
\end{abstract}

\maketitle

\section{Introduction}

Probing shorter distance physics by direct production experiments 
is becoming increasingly difficult.
In the 21-th century, we may have to look for 
small effects at accessible energies in order to learn about
physics at higher energy. If this scenario is correct, 
it is important to develop methods of calculations which can 
outperform the accuracy of the Monte Carlo methods. 

We present
here some results showing that very simple {\it algebraic}
methods can be used to calculate very accurately the renormalized
quantities of Dyson's hierarchical model \cite{dyson}.
The RG transformation of this model is closely related \cite{fam}
to Wilson's approximate recursion formula \cite{wilson}.
In both cases (which are examples of ``hierarchical approximations''), 
the RG transformation maps  
a local potential into another local potential and there is 
no wave function renormalization or generation of higher derivative terms.
Such a simplified 
version of the RG is only a {\it qualitative } approximation
of the one which holds for nearest neighbor interaction lattice models.
However, if we consider the values of the critical exponents,
it is a much better approximation than the gaussian
model. 

As far as the zero momentum Green's functions are concerned,
the numerical
treatment that we proposed in Ref. \cite{finite} is tantamount to a closed
form solution. The method advocated replaces the evaluation
of multiple integrals (which in many lattice models can only be 
performed with the MC methods) 
by the evaluation of a {\it single} integral
followed by purely algebraic manipulations. 
The computer time
involved in this procedure grows only like the logarithm of the
number of sites. These great numerical advantages lead us to
reconsider the question of the improvment of the 
hierarchical approximation. This is a 
hard bookkeeping and computational problem which we plan to 
attack in the future.

Dyson's model couples the main spin in boxes of size
$2^l$ with a strength $({c\over 4})^l$, where $c$ is 
a free parameter set to  
$2^{(D-2)/D}$ in order to approximate
a nearest neighbor scalar model in $D$-dimensions. 
A more systematic presentation of the model can be found for 
instance in Ref. \cite{finite}. 
Models with fermi fields can be constructed similarly
by replacing $D-2$ by $D-1$ in $c$. In addition one needs to specify
a local measure $W_0(\phi)$, for instance of the 
Landau-Ginzburg type ($W_0(\phi) =  e^{-A\phi^2-
B \phi ^4}$) or of the Ising type.
Under a block spin transformation, the local measure changes 
according to:
\begin{eqnarray}
&&W_{n+1}(\phi)\propto e^{{\beta \over 2} ({c\over 4})^{n+1} \phi ^2}\nonumber \\
&&\times\int d\phi' W_n({{(\phi -\phi ')}\over 2})
W_n({{(\phi +\phi ')}\over 2}) \ ,
\end{eqnarray}
This recursion formula can be reexpressed in Fourier representation.
\begin{equation}
R_{n+1}(k)\propto exp(-{1\over 2}\beta 
{{\partial ^2} \over 
{\partial k ^2}})(R_{n}({{\sqrt{c}k}\over 2}))^2  
\end{equation}
where $R_n(k)$ is the Fourier transform of $W_n(\phi)$ with 
an appropriate rescaling \cite{high}. 

\section{The Calculational Method}

Recently, it was found \cite{finite} that finite dimensional
approximations of degree $l_{max}$:
$R_n(k)=1+a_{n,1}k^2 +a_{n,2}k^4+..... +a_{n,l_{max}}k^{2l_{max}}$
provide very accurate results in the symmetric phase and 
moderately accurate results in the broken phase. 
In order to fix the ideas, one 
can calculate the critical exponent $\gamma $ with three decimal
points using $l_{max}=15$ and with 13 decimal points with 
$l_{max}=50$. 
The recursion formula for the $a_{n,m}$ reads
$a_{n+1,m}={u_{n+1,m}\over u_{n+1,0}}$
with
\begin{eqnarray}
u_{n+1,m}&=&{\sum_{l=m}^{l_{max}}(\sum_{p+q=l} a_{n,p}a_{n,q})\Gamma_{l,m}}
\nonumber\\
\Gamma_{l,m}&=&
{{{(2l)!}\over {(l-m)!(2m)!}}
({c\over 4})^l
(-{1\over 2}\beta)^{l-m}}\ .
\label{eq:map}
\end{eqnarray}
The initial condition for the 
Ising measure is
$R_0=cos(k)$. For the Landau-Ginsburg measure, the coefficients in the 
$k-$expansion need to be  evaluated numerically. This is the only 
integral which needs to be calculated, after 
we only have {\it algebraic} manipulations.
The effects of finite dimensional truncations decay faster than 
exponentially. If $\chi^{(l)}$ denotes the susceptibility calculated
with $l_{max}=l$, then \cite{finite}
\begin{equation}
|{{\chi^{(l+1)}-\chi^{(l)}}\over{\chi^l}}|\simeq l^{(-|s|l+q)}
\end{equation}

The volume effects can be controlled arbitrarily well in the symmetric phase
where all the calculations which follow have been performed.
The number $n(\beta , \Delta)$
of iterations necessary to calculate the susceptibility at fixed $\beta$, with
a relative precision 
$\vert{ {\chi_{n+1} -\chi _n}\over {\chi _n}}\vert=\Delta$ can easily be 
estimated as 
\begin{equation}
n(\beta , \Delta )=-({D\over{2}})
(Log _{2}(\Delta)+\gamma Log_{2}
(\beta _c -\beta)) \ .
\end{equation}

The main source of error comes from the round-off errors. Near criticality,
$R_n(k)$ spends many iterations near the fixed point and the errors are 
amplified in the unstable direction. A simple calculation \cite{finite}
shows that if $\delta$ is a typical round-off error (e.g. $10^{-16}$ 
in double precision), then the relative error on the 
susceptibility is of the order $(\beta_c - \beta )^{-1}\delta$. 
Numerical experiments confirm this estimate, however the detailed
statistical properties has some intriguing features
(non-gaussian distributions) which remain to be understood.

\section{Numerical Results}
The method described above can be combined with conventional expansions.
In particular it allows us to calculate several hundred coefficients
of the high-temperature expansion. In $D=3$, it was found \cite{osc} that if 
we use a parametrization of the form
\begin{equation}
\chi=(\beta _c -\beta )^{-\gamma } (A_0 + A_1 (\beta _c -\beta)^{
\Delta } +....)\ 
\label{eq:scaling}
\end{equation}
the $A_0$, $A_1$ are in general log-periodic functions of the form
$\sum_l a_l (\beta_c -\beta)^{ i2\pi l\over log \lambda }$
where $\lambda$ is the largest eigenvalue and $a_1\neq 0$ .
In $D=4$, the logarithmic corrections to the mean-field scaling laws
can also be obtained from the high-temperature expansion. 
Minimizing the errors on $t(m)=((r_m \beta_c -1)m)^{-1}\ $, where $r_m$
is the ratio of two successive HT coefficients, 
for $300\leq m\leq 400$, we found \cite{ht4} that
\begin{equation}
\chi\simeq(\beta _c -\beta )^{-\gamma} (A_0 (|ln(\beta _c -\beta)|)^{p}+A_1)\ 
\end{equation}
with
$\gamma=0.9997$ and
$p=0.3351$. This result is in good agreement with the classic field-theoretical
result $\gamma=1$ and $p=1/3$.

Eq.(\ref{eq:map}) can be used to calculate numerically the renormalized 
coupling constants. Using the notation
$M_n$ for the total field $\sum \phi _x$ inside
blocks of side $2^n$ and $<...>_n $ for 
the average calculated inside these block, we define the 
dimensionless renormalized coupling
\begin{equation}
\lambda_{4,n}= {{{-<M_n^4>_n+3(<M_n^2>_n)^2}}
\over{{2^n}({{<M_n^2>_n} \over {2^n}})^{{D\over2}+2}}}
\end{equation}
The numerator scales like $2^n$ while its two terms scale like  $2^{2n}$
and as $n$ increases, more and more significant digits get lost 
in the subtraction procedure. It is nevertheless possible to 
stabilize \cite{finite} several digits of $\lambda_4$.
We found 
that for $D=3$, $\lambda_4$ reaches a finite
non-zero limit at criticality $\lambda_4^{\ast}=1.92786$ for a Ising measure.
This property is sometimes refered to as hyperscaling. When one approaches 
criticality (which can easily be reformulated as ``when the cut-off becomes
large''), this limiting value is approached according to the approximate law
\begin{equation}
\lambda_4 -\lambda_4^{\ast}\simeq 1.68\times (\beta_c-\beta)^{+0.43} 
\end{equation}
In $D=4$, one can check the ``triviality'' of the theory.
For an Ising measure, 
$\lambda_4$ reaches zero when $\beta$ tends to 
$\beta_c$ according to the approximate law \cite{finite} 
\begin{equation}
(1/\lambda_4 )\simeq {{-1.96-0.746\times ln(\beta_c-\beta)}} 
\end{equation}

One can also use Eq.(\ref{eq:map}) to calculate approximate fixed 
points of the RG transformation. For this purpose, we start with an 
arbitrary measure and we fine-tune $\beta$ until $R_n(k)$ stabilizes 
for a large number of iterations. This can be monitored in terms of 
ratio of successive coefficients. The approximate fixed point 
so obtained are fixed points for a particular value of $\beta$, 
however, it is possible to reabsorb this dependence by a rescaling
of $k$. We have followed this procedure  
for a large class of models \cite{rapid}, namely
$W_0(\phi)\propto exp^{-({1\over 2}m^2 \phi^2+ g\phi^{2p})} $
with
$m^2=\pm 1$ 
(single or double-well potentials), $p=2,3$ or 4 (coupling constants
of positive, zero and negative dimensions when the cut-off is  restored)
and $g=10$ or 0.1 (moderately large and small couplings). All 
the approximate rescaled fixed points
$R^{\star}(\sqrt{\beta_c}k)$ we found turned out to be very close to a
a universal function
\begin{eqnarray}
U(k)=1.- 0.35871*k^2 + 0.05354*k^4
...
\end{eqnarray}
This universal function coincide in very good approximation
with the function which can be obtained from 
the fixed point constructed with great accuracy 
by Koch and Wittwer \cite{wittwer}.
Namely, we found that 
$|\delta u_l|<{{5\times 10^{-5}}\over {l!2^l}}$
where $u_l$ are the coefficients of $U$.
The use of this fixed point allows a very accurate
determination \cite{slow} of the 
critical exponents appearing in Eq. (\ref{eq:scaling}):
$\gamma=1.2991407301586$ and$\Delta=0.4259468589881$. Direct fits of the 
susceptibility confirm 11 decimal points of $\gamma$ and 5 of $\Delta$.

Work in progress involves the improvement of the hierarchical approximation,
development of accurate methods in the broken phase
and tests of perturbation theory. We are also considering hierarchical
Fermi-Bose systems in order to test if it is possible
to construct fully non-perturbative model without fine-tuning. In other words
can the addition of hierarchical fermions with a suitably chosen set of bare couplings to bosons 
drive the boson measure ``naturally'' toward the stable manifold?
Another question being investigated is:
given the fixed point of Dyson model, can we calculate the susceptibility
away from criticality? Or in other words, does the result
of Koch and Wittwer \cite{wittwer} solve the model ?
The answer to this question
seems to be yes in low $l_{max}$ cases.

\end{document}